\documentclass[aps,groupedaddress]{revtex4}
\usepackage{amsfonts}
\usepackage{amsmath}
\usepackage{amssymb}
\usepackage{array}
\usepackage{bm}
\usepackage{braket}
\usepackage{breqn}
\usepackage{color}
\usepackage{dcolumn}
\usepackage{epsfig}
\usepackage{epsf}
\usepackage{epstopdf}
\usepackage{float}
\usepackage{graphicx}
\usepackage{graphics}
\usepackage{harpoon}
\usepackage{indentfirst}
\usepackage{mathrsfs}
\usepackage{multirow}

\newcommand {\sla}[1]{ #1 \!\!\!/}

\begin{document}
\title{Exact Relations for Two-Photon-Exchange Effect in Elastic $ep$ Scattering by Dispersion Relation and Hadronic Model}
\author{
Hui-Yun Cao, Hai-Qing Zhou\protect\footnotemark[1]\protect\footnotetext[1]{E-mail: zhouhq@seu.edu.cn}\\
School of Physics, Southeast University, Nanjing 211189, China\\}
\date{\today}
\begin{abstract}
The two-photon-exchange (TPE) effect plays a key role to extract the form factors (FFs) of the proton. In this work, we discuss some exact properties on the TPE effect in the elastic $ep$ scattering. By taking four low energy interactions as examples, we analyze the kinematical singularities, the asymptotic behaviors and the branch cuts of the TPE amplitudes. The analytic expressions clearly indicate some exact relations between the dispersion relation (DR) method and the hadronic model (HM) method. It suggests that the two methods should be modified to general forms and the new forms give the same results. After the modification the new DRs include a non-trivial term with two singularities. Furthermore, the new DRs automatically include the contributions due to the seagull interaction, the meson-exchange effect, the contact interactions and the off-shell effect. To analyze the elastic $e^{\pm}p$ scattering data sets, the new forms should be used.
\end{abstract}
\maketitle
\section{Introduction}
The proton is the unique stable hadron and one of the elemental constituents of our world. The knowledge on its properties is an important base to understand the world. In the last twenty years our knowledge on its structure has been improved greatly but some puzzles still exist. The electromagnetic form factors (EM FFs) of the proton are two of the most elemental and well-defined non-perturbative quantities reflecting its structures. To extract the EM FFs of the proton, the precise experimental data sets of the elastic $e^{\pm}p$ or $\mu p$ scattering \cite{Ex-precise-measurements} are necessary. To analyze and understand these precise experimental data sets the two-photon-exchange (TPE) effect is the key. From 2003 to now, many theoretical dynamical methods and model independent analysis are suggested and applied to estimate the TPE contribution such as the hadronic model (HM) method \cite{hadronic-model-method}, the GPDs method \cite{GPD-method}, the dispersion relation (DR) method \cite{dispersion-relation-11,dispersion-relation-12,dispersion-relation-21,dispersion-relation-22}, the perturbative QCD \cite{pQCD-method}, the soft collinear effective theory \cite{SCEF-method}, the chiral perturbative theory \cite{TPE-chiral}, and the parametrization method \cite{phenomenological-parametrizations}. But it is still far away from the accurate estimation of the TPE contribution.

To analyze the experimental data sets below a few GeV$^2$, the HM method and the DR method are usually used. In the HM method, the interactions between the photon and the intermediate states (such as nucleon and $\Delta(1232)$) are constructed to estimate the TPE amplitude manifestly. In the DR method only the interactions between the photon and the on-shell intermediate states (such as nucleon, $\Delta(1232)$ and $\pi N$ continue state) are used to estimate the imaginary part of the TPE amplitude in the physical region. After analytically continuing the imaginary part of the TPE amplitude to the unphysical region and combining it with the asymptotic behavior of the TPE amplitude, the real part of the TPE amplitude can be got. This means that the DR method only uses the on-shell FFs. It is often argued that this is a big goodness of the DR method comparing with the HM method since the latter may include the off-shell information. Another goodness of the DR method is the good behavior of the TPE contribution in the Regge limit. The HM method results in un-physical behavior in the Regge limit when the excited intermediate states (such as  $\Delta(1232)$) are considered. Due to these advantages, recently the DR method is widely accepted and applied to analyze the experimental data sets \cite{dispersion-relation-12,dispersion-relation-21,dispersion-relation-22}. But this does not mean that the DR method is a perfect and uniquely reliable method. For example, two different DRs are used in Ref. \cite{dispersion-relation-11,dispersion-relation-12} and Ref. \cite{dispersion-relation-21}, and the contributions from the meson-exchange effect \cite{Zhouhq-Meson-exchange-2014,Afanasev-Meson-exchange,Borisyuk-Meson-exchange,Zhouhq-Meson-exchange-2017,Dorokhov-Meson-exchange-2018} are not included.

In this work, at first we analyze the analytic structures of the TPE amplitudes in four typical and general toy interactions. Since these interactions are valid at low energy, then we use them to check low energy's behaviors of the TPE amplitudes by the DR method. The detailed comparison and analysis clearly show some important properties on the TPE contributions by the DR method and the HM method: (1) the DR method and the HM method used in the references should be modified to general forms to correctly include the contributions from the seagull interaction, the meson-exchange effect, the contact interactions and the off-shell effect; (2) after the modification the new DRs include a non-trivial term with two singularities; (3) the two methods after the modification give the exact same results.

\section{Basic Formula}
In the limit $m_e\rightarrow0$, the amplitude of the elastic $ep$ scattering  with $C,P,T$ invariance can be written as
\begin{eqnarray}
{\cal M}(ep\rightarrow ep)&=&\sum_{i=1}^{3} {\cal F}_i{\cal M}_i,
\end{eqnarray}
with
\begin{eqnarray}
{\cal M}_{1} &\overset{def}{=}& M_N[\overline{u}_3\gamma_{\mu}u_1][\overline{u}_4\gamma^{\mu} u_2], \nonumber\\
{\cal M}_{2} &\overset{def}{=}& [\overline{u}_3(\sla{p}_2+\sla{p}_4)u_1][\overline{u}_4u_2], \nonumber\\
{\cal M}_{3} &\overset{def}{=}& M_N[\overline{u}_3\gamma_5\gamma_{\mu} u_1][ \overline{u}_4\gamma_5\gamma^{\mu} u_2],
\end{eqnarray}
where we shortly write $\overline{u}(p_i,m_i,h_i)$ and $u(p_i,m_i,h_i)$ as $\overline{u}_i$ and $u_i$, $p_{1,2}$ are the momenta of the initial electron and proton, $p_{3,4}$ are the momenta of the final electron and proton, $h_i$ are the helicities of the corresponding spinors, $m_{1,3}=m_e$ and $m_{2,4}=M_N$.

One can calculate the coefficients ${\cal F}_i$ by solving the following algebraic equations:
\begin{eqnarray}
\sum_{helicity}{\cal M}{\cal M}^{*}_j =\sum_{i=1}^{3} \sum_{helicity} {\cal F}_i{\cal M}_i{\cal M}^{*}_j.
\end{eqnarray}
After some simple calculation, the coefficients ${\cal F}_i$ can be expressed as
\begin{eqnarray}
{\cal F}_{i}=\sum_{j}({\cal D}^{-1})_{ij} \sum_{helicity}{\cal M}{\cal M}^*_j,
\label{Eq:project-result}
\end{eqnarray}
with ${\cal D}_{ij}\overset{def}{=}\sum{\cal M}_{i}{\cal M}^*_{j}$. In four dimension the matrix ${\cal D}^{-1}$  is expressed as
\begin{eqnarray}
{\cal D}^{-1}&=&\frac{1}{-4M_N^2t(\nu^2-\nu_s^2)^2}
\begin{pmatrix}
\overline{d}_{11} & \overline{d}_{12} & \overline{d}_{13}\\
\overline{d}_{21} & \overline{d}_{22} & \overline{d}_{23}\\
\overline{d}_{31} & \overline{d}_{32} & \overline{d}_{33} \\
\end{pmatrix},
\label{Eq:inverse-D}
\end{eqnarray}
with
\begin{eqnarray}
\overline{d}_{11} &=& (4M_N^2-t)(\nu^2+\nu_s^2),  \nonumber\\
\overline{d}_{22} &=& M_N^2(\nu^2-t(4M_N^2+t)), \nonumber\\
\overline{d}_{33} &=& -t(\nu^2+\nu_s^2),\nonumber\\
\overline{d}_{12} &=& \overline{d}_{21}=-2M_N^2(\nu^2+\nu_s^2),\nonumber\\
\overline{d}_{13} &=& \overline{d}_{31}=2t(4M_N^2-t)\nu,\nonumber\\
\overline{d}_{23} &=& \overline{d}_{32}=-4M_N^2t\nu.
\end{eqnarray}
where $\nu\overset{def}{=}(p_1+p_3)\cdot(p_2+p_4)$, $t\overset{def}{=}(p_1-p_3)^2$, and $\nu_s\overset{def}{=}\sqrt{-t(4M_N^2-t)}$. One can see that ${\cal D}^{-1}$  has two kinematic singularities in the unphysical region when $\nu\rightarrow\pm \nu_s$. The similar singularities have been discussed in  Ref. \cite{dispersion-relation-12} and are directly neglected when applying the DRs. These kinematic singularities are not physical poles but related with the definition of ${\cal F}_i$ in the physical region. Usually such kinematic singularities are canceled by the corresponding factor in $\sum{\cal M}{\cal M}^*_j$, while in some special cases this cancellation does not happen and we show this in the following.

In the one-photon exchange (OPE) approximation, the amplitude ${\cal M}^{(1\gamma)}$ and the corresponding coefficients ${\cal F}_{i}^{(1\gamma)}$ can be got easily. The final results ${\cal F}_i^{(1\gamma)}$ are free from the kinematic singularities, only dependent on $t$ and their imaginary parts are exact zero.

\begin{figure}[bhtp]
\centering
\includegraphics[height=7cm]{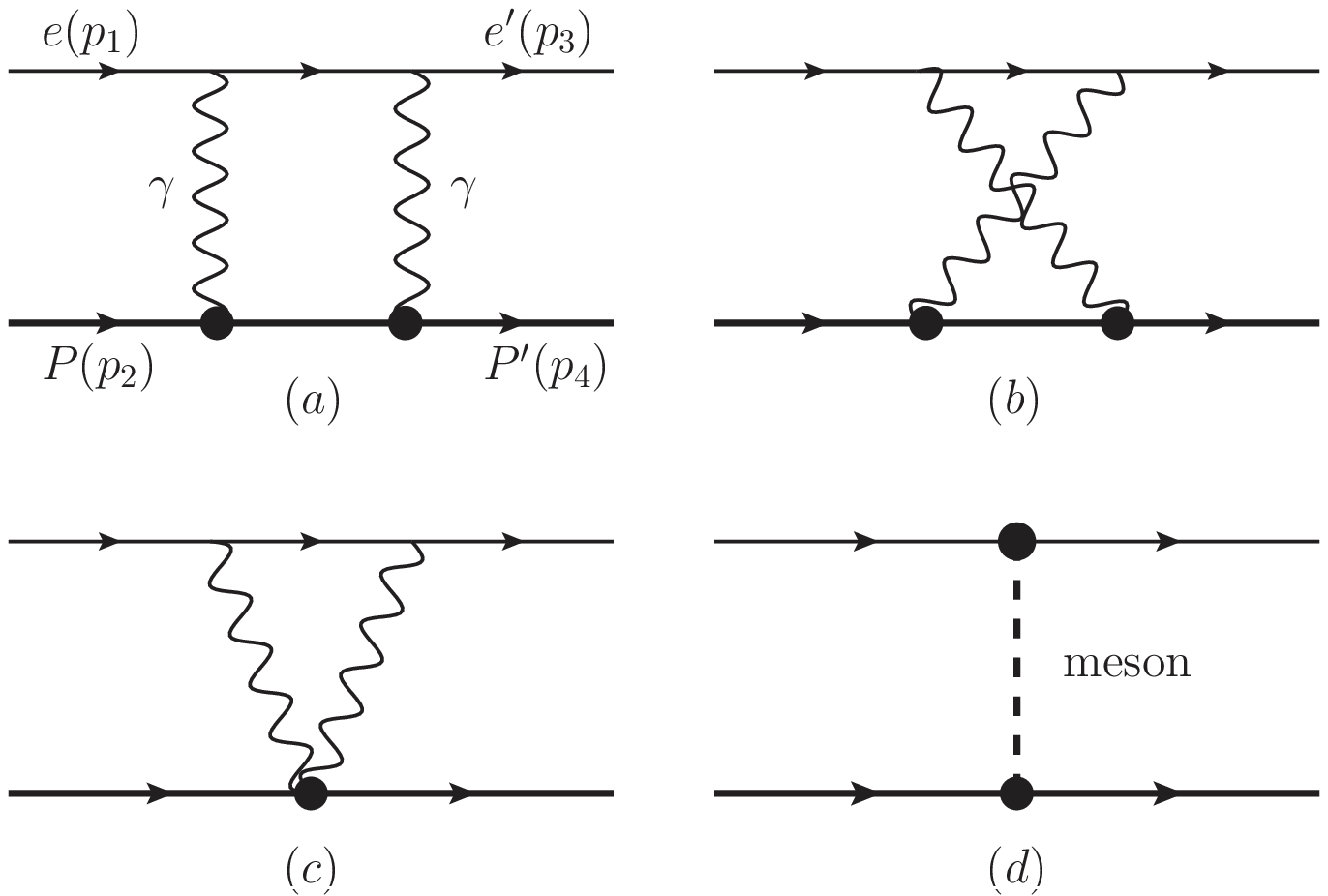}
\caption{The possible two-photon-exchange effects in elastic $ep$ scattering where only the nucleon intermediate state is considered with $(a)$ the box diagram, $(b)$ the crossed-box diagram, $(c)$ the seagull diagram and $(d)$ the meson-exchange diagram.}
\label{Figure:TPE-four-type}
\end{figure}

\begin{figure}[bhtp]
\centering
\includegraphics[height=3cm]{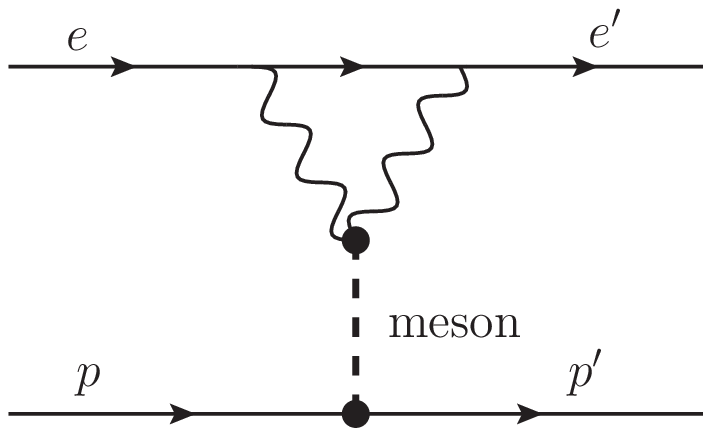}
\caption{The possible two-photon-exchange effects in elastic $ep$ scattering where only the nucleon intermediate state is considered with $(a)$ the box diagram, $(b)$ the crossed-box diagram, $(c)$ the seagull diagram and $(d)$ the meson-exchange diagram.}
\label{Figure:meson-exchange-loop}
\end{figure}

When go beyond the OPE approximation, the TPE effect should be considered. In this work we do not consider the TPE contributions from the baryon resonances and the $\pi N$ continue states since in principle their contributions can be dealt similarly. In this case, dynamically there are four types of contributions showed in Fig. \ref{Figure:TPE-four-type} where $(a,b)$ are the usual box and crossed-box diagrams, $(c)$ is the seagull diagram  and $(d)$ refers to the meson-exchange effect. For the meson-exchange effect, in principle one should consider the coupling between the mesons and two photons like the triangle diagram Fig. \ref{Figure:meson-exchange-loop} which are discussed in Ref.\cite{Zhouhq-Meson-exchange-2014,Afanasev-Meson-exchange,Borisyuk-Meson-exchange,Zhouhq-Meson-exchange-2017,Dorokhov-Meson-exchange-2018}. While after the loop integration the final results can be expressed in a general form like Fig. (\ref{Figure:TPE-four-type}$d$). Here for simplicity we directly take Fig. (\ref{Figure:TPE-four-type}$d$) as example to discuss the analytic TPE property.

Due to the crossing symmetry, one has the following general relations when $t<0$:
\begin{eqnarray}
{\cal F}^{(a,c,d)}_{1,2}(t,\nu^+)&=&-{\cal F}^{(b,c,d)}_{1,2}(t,-\nu^+),\nonumber\\
{\cal F}^{(a,c,d)}_{3}(t,\nu^+) &=&{\cal F}^{(b,c,d)}_{3}(t,-\nu^+),
\label{Eq:relation-abcd}
\end{eqnarray}
where we use ${\cal F}_i(t,\nu)$ to refer to the coefficients of the TPE amplitude, $\overline{\nu}^+=\overline{\nu}+i0^+$, and the indexes $(a,b,c,d)$ are corresponding to the diagrams $(a,b,c,d)$ in Fig. \ref{Figure:TPE-four-type}.

In Ref. \cite{dispersion-relation-11,dispersion-relation-12} the following DRs are used to estimate the physical TPE contributions:
\begin{eqnarray}
\textrm{Re}[{\cal F}_{1,2}^{\textrm{DR1}}(t,\nu)]
&=&\frac{2\nu}{\pi}\textrm{P}\Big[\int_{\nu_{th}}^{\infty}\frac{\textrm{Im}[{\cal F}^{(a)}_{1,2}(t,\overline{\nu}^+)]}
{\overline{\nu}^2-\nu^2}d\overline{\nu}\Big], \nonumber\\
\textrm{Re}[{\cal F}_{3}^{\textrm{DR1}}(t,\nu)]
&=&\frac{2}{\pi}\textrm{P}\Big[\int_{\nu_{th}}^{\infty}\frac{\overline{\nu}\textrm{Im}[{\cal F}^{(a)}_{3}(t,\overline{\nu}^+)]}
{\overline{\nu}^2-\nu^2}d\overline{\nu}\Big],
\label{Eq:DR1}
\end{eqnarray}
where the index DR1 refers to the method used in Ref. \cite{dispersion-relation-11,dispersion-relation-12}, the operator $\textrm{P}$ refers to the principle value integration and $\nu_{th}=t$. In Ref. \cite{dispersion-relation-21} the DRs are modified as follows:
\begin{eqnarray}
\textrm{Re}[{\cal F}_{1,2}^{\textrm{DR}2}(t,\nu)]
&=&\textrm{Re}[{\cal F}_{1,2}^{\textrm{DR}1}(t,\nu)],\nonumber\\
\textrm{Re}[{\cal F}_{3}^{\textrm{DR}2}(t,\nu)]
&=&\textrm{Re}[{\cal F}_{3}^{\textrm{DR2}}(t,\nu_0)]+\frac{2(\nu^2-\nu_0^2)}{\pi}\textrm{P}\Big[\int_{\nu_{th}}^{\infty}\frac{\overline{\nu}\textrm{Im}[{\cal F}^{(a)}_{3}(t,\overline{\nu}^+)]}
{(\overline{\nu}^2-\nu^2)(\overline{\nu}^2-\nu_0^2)}d\overline{\nu}\Big],
\label{Eq:DR2}
\end{eqnarray}
where the index DR2 refers to the method used in Ref. \cite{dispersion-relation-21}, $\nu_0$ is any fixed number, $\textrm{Re}[{\cal F}_{3}^{\textrm{DR}2}(t,\nu_0)]$ is an unknown function which can be determined by the experimental data sets at fixed $\nu_0$, and the final result is not dependent on $\nu_0$.

These DRs are widely accepted to replace the HM method to estimate the TPE contributions and to analyze the experimental data sets. While naively one can easily check that these DRs do not include the contributions from Fig.\ref{Figure:TPE-four-type}($c,d$) since their imaginary parts are exact zero when $t<0$. To understand and solve this problem, in this work at first we take the following four low energy interactions as examples to show the analytic properties of the TPE amplitudes in these interactions:
\begin{eqnarray}
{\cal L}_{E} &\overset{def}{=}& -e \overline{\psi}_p\gamma^\mu\psi_p A_\mu,\nonumber\\
{\cal L}_{M} &\overset{def}{=}& -\frac{e\kappa}{4M_N} \overline{\psi}_p\sigma^{\mu\nu}\psi_p F_{\mu\nu},\nonumber\\
{\cal L}_{S}&\overset{def}{=}& -\frac{2\pi}{M_N^2}(\partial_{\mu}\overline{\psi}_p)(\partial_{\nu}\psi_p)
(\alpha_{E1}F^{\mu\rho}{F}^{\nu}_{\rho}+\beta_{M1}\widetilde{F}^{\mu\rho}\widetilde{F}^{\nu}_{\rho}),\nonumber\\
{\cal L}_{T} &\overset{def}{=}& ig_{\textrm{Tpp}}[(\partial_\mu\overline{\psi}_p)\gamma_\nu\psi_p-\overline{\psi}_p\gamma_\nu(\partial_\mu\psi_p) ]\phi^{\mu\nu}+ig_{\textrm{Tee}}[(\partial_\mu\overline{\psi}_e)\gamma_\nu\psi_e-\overline{\psi}_e\gamma_\nu(\partial_\mu\psi_e)]  \phi^{\mu\nu},
\end{eqnarray}
where $\psi_{p},A_\mu,\psi_e$ and $\phi_{\mu\nu}$ refer to the fields of proton, photon, electron and tensor meson, respectively. Similar with the reason of Fig. (\ref{Figure:TPE-four-type}$d$), one can take the direct coupling between the meson and the electrons to discuss the behavior of the TPE amplitude since the loop integration does not change the analytic property in the region with $t<0$. Furthermore, here we only take $2^{++}$ meson as example to show the analytic property since the contributions from the mesons with other quantum numbers are similar when $t$ is fixed as negative. 

By these interactions, the corresponding amplitudes ${\cal M}_{E,M}^{(a,b)},{\cal M}_{S}^{(c)},{\cal M}_{T}^{(d)}$ and ${\cal F}_{Xi}^{(y)}(t,\nu)$ can be got easily, where $X$ refers to $E,M,S,T$ and $y$ refers to $a,b,c,d$, respectively, and
\begin{eqnarray}
{\cal M}_{X}^{(y)} \overset{def}{=} \sum_{i=1}^{3} {\cal F}_{Xi}^{(y)}(t,\nu){\cal M}_i.
\end{eqnarray}

\section{Analytic results and discussion}
In the practical calculation, at first we use Eq. (\ref{Eq:project-result}) to get the expressions of the coefficients ${\cal F}_{Xi}^{(y)}(t,\nu)$ in $d$ dimension, then do the loop integration with the dimensional regularization. The packages FeynCalc \cite{FenyCalc} and PackageX \cite{Pacakge X} are used to do the analytic calculation. After the loop integration, we expand ${\cal F}_{Xi}^{(y)}(t,\nu)$ at $\nu=\pm \nu_s$ to analyze the kinematic singularities and expand them at $\nu=\pm\infty$ to get their asymptotic behaviors. The imaginary parts and the discontinuities of ${\cal F}_{Xi}^{(y)}(t,\nu)$ in the complex plane of $\nu$ and $t$ are used to analyze the branch cuts.

The final analytic results show that the kinematical singularities are cancelled in ${\cal L}_{E,S,T}$ cases but are remained in ${\cal L}_{M}$ case.  The analytic asymptotic behaviors of the coefficients are expressed as follows:
\begin{eqnarray}
\textrm{Re}[{\cal F}^{(a)}_{E1}(t,\nu)] &\overset{\nu\rightarrow \infty}{\longrightarrow}& -\frac{4\alpha_e^2}{M_Nt}[(\frac{1}{\widetilde{\epsilon}_{\textrm{IR}}}+\ln\frac{\overline{\mu}^2_{\textrm{IR}}}{-t})\ln {\nu}+c_{1}],\nonumber\\
\textrm{Im}[{\cal F}^{(a)}_{E1}(t,\nu^+)] &\overset{\nu\rightarrow \infty}{\longrightarrow}& \frac{4\pi\alpha_e^2}{M_Nt}(\frac{1}{\widetilde{\epsilon}_{\textrm{IR}}}+\ln\frac{\overline{\mu}_{\textrm{IR}}^2}{-t}), \nonumber\\
{\cal F}_{E2,E3}^{(a)}(t,\nu) &\overset{\nu\rightarrow \infty}{\longrightarrow}& 0,
\end{eqnarray}
\begin{eqnarray}
\textrm{Re}[{\cal F}^{(a)}_{M1}(t,\nu)] &\overset{\nu\rightarrow \infty}{\longrightarrow}& -\frac{\alpha_e^2\kappa^2}{4M_N^3}\Big[\ln^2
\nu-2(1+\ln(-2t))\ln\nu +c_{2}+\frac{3}{4}\frac{1}{\widetilde{\epsilon}_{\textrm{UV}}}\Big],\nonumber\\
\textrm{Re}[{\cal F}^{(a)}_{M2}(t,\nu)] &\overset{\nu\rightarrow \infty}{\longrightarrow}& \frac{\alpha_e^2\kappa^2}{4M_N^3}c_{3}\nonumber\\
\textrm{Re}[{\cal F}^{(a)}_{M3}(t,\nu)] &\overset{\nu\rightarrow \infty}{\longrightarrow}&
  -\frac{\alpha_e^2\kappa^2}{8M_N^3}\Big(5+3\ln\frac{\overline{\mu}_{\textrm{UV}}^2}{-t}+\frac{3}{\widetilde{\epsilon}_{\textrm{UV}}}\Big),\nonumber\\
\textrm{Im}[{\cal F}^{(a)}_{M1}(t,\nu^+)] &\overset{\nu\rightarrow \infty}{\longrightarrow}& \frac{\pi\alpha_e^2 \kappa^2}{2M_N^3}\Big[\ln \frac{\nu}{-t}-(1+\ln2)\Big], \nonumber\\
\textrm{Im}[{\cal F}_{M2,M3}^{(a)}(t,\nu^+)] &\overset{\nu\rightarrow \infty}{\longrightarrow}& 0,
\end{eqnarray}
\begin{eqnarray}
\textrm{Re}[{\cal F}^{(c)}_{S2}(t,\nu)] &=&  -\frac{\alpha_e (\alpha_{E1}+\beta_{E1})}{72M_N^2}\Big(17+12\log\frac{\overline{\mu}_{\textrm{UV}}^2}{-t}+12\frac{1}{\widetilde{\epsilon}_{\textrm{UV}}}\Big)\nu,\nonumber\\
\textrm{Re}[{\cal F}^{(d)}_{T1}(t,\nu)] &=& \frac{g_{\textrm{Tee}} g_{\textrm{Tpp}}}{M_N(M_T^2-t)}\nu,\nonumber\\
\textrm{Re}[{\cal F}^{(d)}_{T3}(t,\nu)] &=& \frac{g_{\textrm{Tee}} g_{\textrm{Tpp}}}{2M_{N}(M_T^2-t)}t,\nonumber\\
{\cal F}_{S1,S3,T2}^{(c,d)}(t,\nu) &=& 0, \nonumber\\
\textrm{Im}{\cal F}_{S,T}^{(c,d)}(t,\nu) &=& 0,
\label{Eq:TPE-Tensor}
\end{eqnarray}
where $\alpha_e=e^2/4\pi$, $\overline{\mu}_{\textrm{IR,UV}}$ are the $\textrm{IR}$ and $\textrm{UV}$ scales, $c_i$ are some simple functions independent on $\nu$ and we do not list them here, and
\begin{eqnarray}
\frac{1}{\widetilde{\epsilon}_{\textrm{IR,UV}}} &=& \frac{1}{\epsilon_{\textrm{IR,UV}}}-\gamma_{E}+\ln4\pi. \nonumber
\end{eqnarray}
The asymptotical behaviors of ${\cal F}^{(b)}_{Xi}(t,\nu)$ can be got easily via Eq. (\ref{Eq:relation-abcd}).

\begin{figure}[htbp]
\center{\epsfxsize 2.8 truein\epsfbox{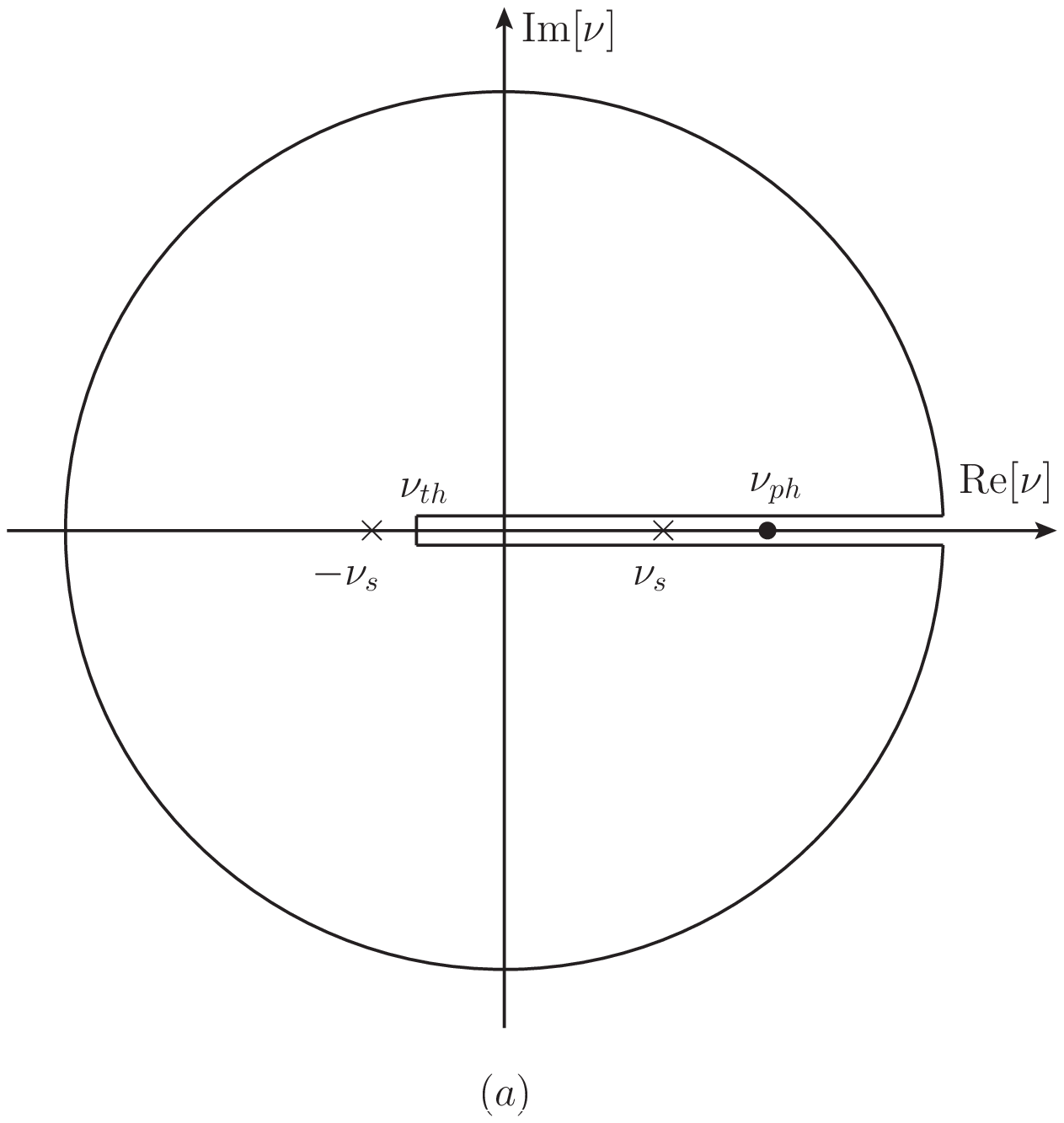}\epsfxsize 2.8 truein\epsfbox{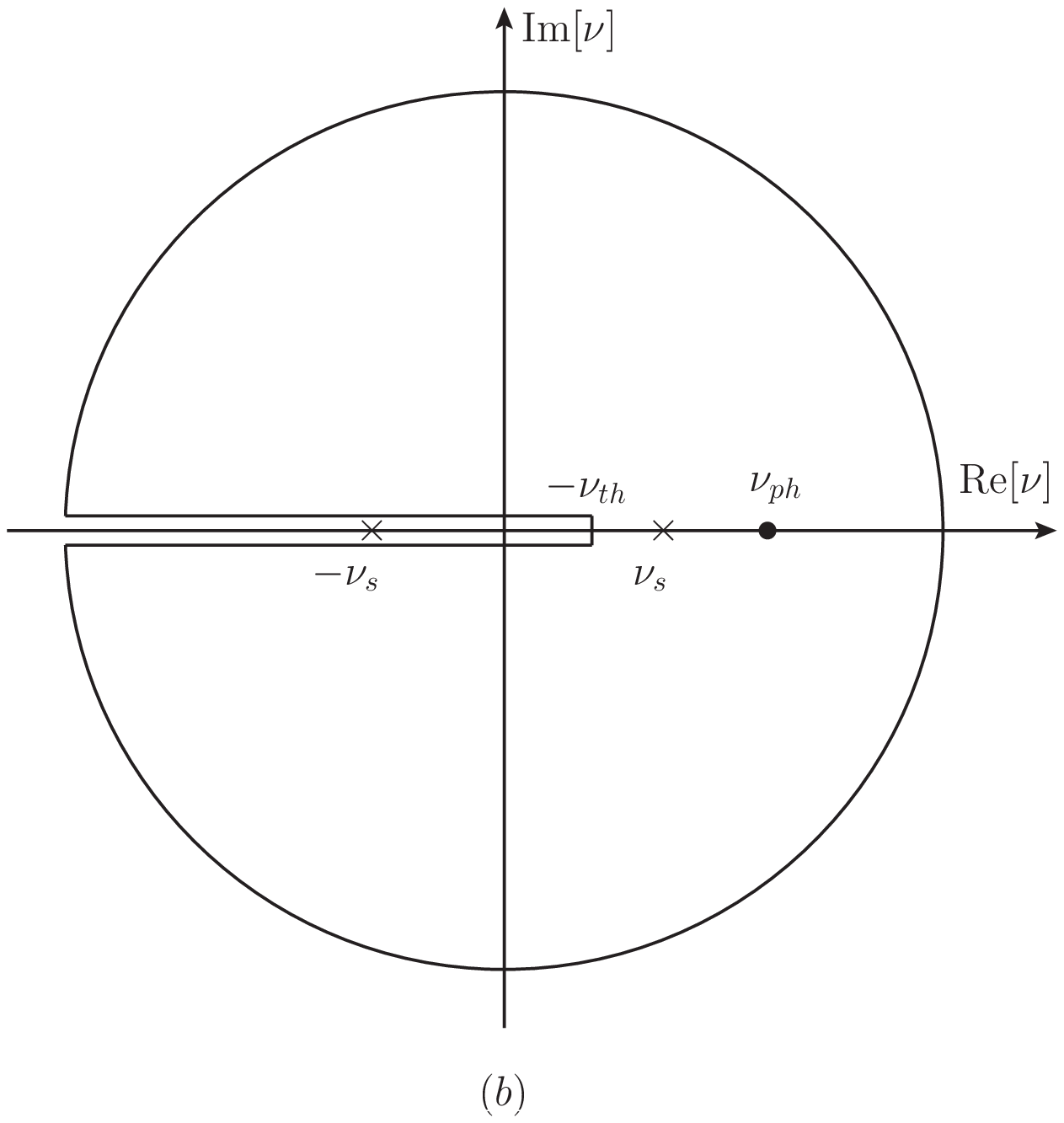}}
\caption{The branch cuts of ${\cal F}_i^{(a,b)}(Q^2,\nu)$ in the complex plane of $\nu$ at fixed negtive $t$.  (a) is for ${\cal F}_i^{(a)}(t,\nu)$ and (b) is for ${\cal F}_i^{(b)}(t,\nu)$ where the physical $\nu_{ph}$ and the possible kinematical singularities are also marked.}
\label{Figure:Branch-cuts}
\end{figure}

On the branch cuts, the analytic results show: (1) when $t<0$, ${\cal F}_{Xi}^{(a)}(t,\nu)$ have one right hand branch cut in the region $\nu\subset [\nu_{th},\infty]$, ${\cal F}_{Xi}^{(b)}(t,\nu)$ have one left hand branch cut in the region $\nu\subset [-\infty,-\nu_{th}]$  and ${\cal F}_{Xi}^{(c,d)}(t,\nu)$ have no branch cut. There properties are  showed in Fig. \ref{Figure:Branch-cuts} and are natural due to the unitarity usually argued in the references. (2) when $t>0$, ${\cal F}_{Xi}^{(a,b,c,d)}(t,\nu)$ have an additional branch cut at real axis of $t$ and this discontinuity on $t$ results in the nonzero imaginary parts of ${\cal F}_{Xi}^{(a,b,c,d)}(t,\nu)$. This is also natural since when $t>0$ the coefficients ${\cal F}_{Xi}^{(y)}(t,\nu)$ are related with the TPE contributions in $e^+e^-\rightarrow p\overline{p}$.

Combing the above properties it is easy to check that ${\cal F}_{Ei}^{(a+b)}(t,\nu)$ satisfy  Eq. (\ref{Eq:DR1}) and ${\cal F}_{Mi}^{(a+b)}(t,\nu)-{\cal F}_{Mi,\textrm{ks}}^{(a+b)}(t,\nu)$ satisfy Eq. (\ref{Eq:DR2}) when $t<0$, where ${\cal F}_{Mi,\textrm{ks}}^{(a+b)}(t,\nu)$ include two kinematic singularities and are expressed as

\begin{eqnarray}
{\cal F}_{Mi,\textrm{ks}}^{(a+b)}(t,\nu)=\frac{A^{(a+b)}_i(\nu,t)}{(\nu^2-\nu_s^2)^2},
\label{Eq:contribution-with-singulariy}
\end{eqnarray}
with $A^{(a+b)}_i(\nu,t)$ being three polynomials of $\nu$ and $t$, their manifest expressions are a little long and are listed in the appendix.
Furthermore, ${\cal F}_{M3}^{(a+b)}(t,\nu)$ includes an UV divergence which means that the single interaction ${\cal L}_{M}$ is not consistent and the corresponding contact interactions should be considered to absorb the UV divergence. This means that the usual HM method should be modified to include the contact interactions. This is the reason that the DR2 is used in Ref. \cite{dispersion-relation-21} to replace the DR1.

We also find if one gives the two photons some virtual masses (different or same) in the Feynman gauge, the terms ${\cal F}_{Mi,\textrm{ks}}^{(a+b)}(t,\nu)$ are not changed. This means that the contributions with the kinematical singularities are exactly cancelled when one replaces the vertex $\Gamma_{M}^{\mu}(k)$ (from ${\cal L}_M$) by $\Gamma_{M}^{\mu}(k)F(k^2)$ with $F(k^2)$ being a monopole like FF  as follow:
\begin{eqnarray}
F(k) &=& \sum_{j}\frac{d_j}{(k^2-\Lambda^2_j)^{n_j}},
\end{eqnarray}
where $k$ is the momentum of the incoming photon, $n_j$ are some natural numbers, and $d_j,\Lambda_j$ are some real parameters.
This cancellation is due to the following simple relation and its generalization.
\begin{eqnarray}
\frac{B}{(k^2-z_1^2)(k^2-z_2^2)} &=& \frac{1}{z_1^2-z_2^2}\big[\frac{B}{k^2-z_1^2}-\frac{B}{k^2-z_2^2}\big].\nonumber
\end{eqnarray}
Since ${\cal F}_{Mi,\textrm{ks}}^{(a+b)}(t,\nu)$ are not dependent on the parameters $z_j$ then the two contributions with the kinematical singularities are cancelled. This property means that the coefficients ${\cal F}_{i}^{(a+b)}(t,\nu)$ in the usual HM method with monopole like FFs as inputs are free from any kinematical singularities. This explains the numerical property of Fig. 19 in Ref. \cite{dispersion-relation-21} where the difference between the DR method and the HM method with FFs are presented. The important point is that this does mean that the kinematical singularities are canceled certainly in any cases.

Another important property is that $\textrm{Im}[{\cal F}_{Si}^{(c)}(t,\nu)]$ and $\textrm{Im}[{\cal F}_{Ti}^{(d)}(t,\nu)]$ are exact zero, but $\textrm{Re}[{\cal F}_{S2}^{(c)}(t,\nu)]$ and $\textrm{Re}[{\cal F}_{T1,T3}^{(d)}(t,\nu)]$ are not zero and satisfy twice-subtracted and once-subtracted DRs, respectively. Similarly there is an UV divergence in $\textrm{Re}[{\cal F}_{S2}^{(c)}]$ which means the contact interactions should be included to absorb the UV divergence. These properties are general when extending the interactions to general forms by including more derivatives. Similarly when the mesons with other $J$ are considered the results are still polynomial functions on $\nu$. After combing these contributions, one can see that the contributions from the seagull interactions, the meson-exchange, and contact interaction can be expressed as
\begin{eqnarray}
{\cal F}_{1,2}^{(c+d)}(t,\nu)] &=&\sum_{j=0}c_{1j,2j}(t)\nu^{2j+1},\nonumber\\
{\cal F}_{3}^{(c+d)}(t,\nu)] &=&\sum_{j=0}c_{3j}(t)\nu^{2j},
\label{Eq:contributions-cd}
\end{eqnarray}
where the properties Eq. (\ref{Eq:relation-abcd}) have been used. At first glance, these results are unphysical  at high energy and are directly neglected in the usual calculation. But their physical meaning can be seen clearly when continue the results to the physical region of $e^+e^-\rightarrow p\overline{p}$ where the variable $\nu$ is corresponding to $-\nu_s\cos\theta_p$ with $\theta_p$ the angle of the final proton's three momentum in the center of mass frame. The physical regions of $\nu,t$ in $ep\rightarrow ep$ and $e^+e^-\rightarrow p\overline{p}$ mean that the contributions Eq. (\ref{Eq:contributions-cd}) are converge in the regions $\nu\rightarrow \infty$ and $|\nu|<\nu_s$.  This means one can express them as follows:
\begin{eqnarray}
\sum_{j=0}c_{1j,2j}(t)\nu^{2j+1}&=&\sum_{j=1}\frac{g_{1j,2j}(t)\nu}{(\nu^2-\nu_s^2)^j}\approx \frac{g_{11,21}(t)\nu}{\nu^2-\nu_s^2},\nonumber\\
\sum_{j=0}c_{3j}(t)\nu^{2j}&=&\sum_{j=0}\frac{g_{3j}(t)}{(\nu^2-\nu_s^2)^j}\approx g_{30}(t),
\label{Eq:contributions-cd-continue}
\end{eqnarray}
where $g_{ij}(t)$ are unknown functions and only the leading contributions are kept since $\nu^2-\nu_s^2$ increase quickly when $\nu$ increase at fixed $t$. The interesting property is that these contributions are the same with Eq. (\ref{Eq:contribution-with-singulariy}) in the leading order.

Finally, one get the following DRs in the leading order of $M_N^4/(\nu^2-\nu_s^2)$:
\begin{eqnarray}
\textrm{Re}[{\cal F}_{1,2}^{\textrm{DR3}}(t,\nu)]
&=&\frac{f_{1,2}(t) \nu}{\nu^2-\nu_s^2}+\frac{2\nu}{\pi}\textrm{P}\Big[\int_{\nu_{th}}^{\infty}\frac{\textrm{Im}[{\cal F}^{(a)}_{1,2}(t,\overline{\nu}^+)]}
{\overline{\nu}^2-\nu^2}d\overline{\nu}\Big],\nonumber\\
\textrm{Re}[{\cal F}_{3}^{\textrm{DR3}}(t,\nu)]
&=&f_3(t)+\frac{2(\nu^2-\nu_0^2)}{\pi}\textrm{P}\Big[\int_{\nu_{th}}^{\infty}\frac{\overline{\nu}\textrm{Im}[{\cal F}^{(a)}_{3}(t,\overline{\nu}^+)]}
{(\overline{\nu}^2-\nu^2)(\overline{\nu}^2-\nu_0^2)}d\overline{\nu}\Big]
\label{Eq:DR-final}
\end{eqnarray}
where we use $f_i(t)$ to refer to the unknown functions and $f_3(t)=\textrm{Re}[{\cal F}_{3}^{\textrm{DR3}}(t,\nu_0)]$.
Eqs. (\ref{Eq:contributions-cd},\ref{Eq:contributions-cd-continue},\ref{Eq:DR-final}) mean the following exact relations between the DR3 and the modified HM method:
\begin{eqnarray}
{\cal F}^{\textrm{DR3}}_{1,2}(t,\nu)&=& \Big[{\cal F}^{(a+b)}_{1,2}(t,\nu)+\sum_{j=0}h_{1j,2j}(t)\nu^{2j+1}\Big]_{\textrm{Ana+LO}},\nonumber\\
{\cal F}^{\textrm{DR3}}_{3}(t,\nu) &=& \Big[{\cal F}^{(a+b)}_{3}(t,\nu)+\sum_{j=0}h_{3j}(t)\nu^{2j}\Big]_{\textrm{Ana+LO}},
\label{Eq:relation-HM-DR}
\end{eqnarray}
where the subindex $\textrm{Ana+LO}$ refers to do analytic continue and keep the leading order contribution, $h_{ij}(t)$ include the contributions from the seagull interaction, the meson-exchange effect and the contact interactions.

Actually, the contributions due to the off-shell effects in $(a+b)$ can also be expressed by $h_{ij}(t)$. This can be understood in a direct physical way. In the DR method, only the on-shell vertex of $\gamma^*NN$ is used to estimate the imaginary parts of the coefficients and the real parts are got by the DRs. In the HM method, if the off-shell vertex is used, one can separate the vertex into two parts as
\begin{eqnarray}
\Gamma_\mu^{\textrm{off-shell}}(p_{i,f}^2,k^2)=\Gamma_\mu^{\textrm{on-shell}}(k^2)+\Delta \Gamma(p_{i,f}^2-M_N^2,k^2),\nonumber
\end{eqnarray}
where  $p_i,p_f$ are the momenta of the initial and final proton in the vertex, respectively,  $\Delta \Gamma(p_{i,f}^2-M_N^2,k^2)$ is a polynomial function on $p_i^2-M_N^2$ or $p_f^2-M_N^2$ when no additional phenomenological poles on $p_i^2$ and $p_f^2$ are introduced in the vertex.
Then the TPE amplitude can be separated into two parts: one only includes the on-shell information and another includes off-shell effect. Naively the first one is just what the DR1 method gives. The second one has a global factor like $p_i^2-M_N^2$ or $p_f^2-M_N^2$ in the numerator. This factor cancels the denominator of the nucleon's propagator and the final result after the loop integration is similar with the contribution from the seagull interaction. This means that the contributions due to the off-shell effect can be expressed by some polynomials on $\nu$ like the contributions from the seagull interaction, the meson-exchange effect and the contact interactions. This property clearly indicates the physical meaning of $h_{ij}(t)$. They include all the contributions due to the seagull interaction, the meson-exchange effect, the contact interactions and the off-shell effect.

\section{Summary}

In summary, the detailed analysis based on four typical and general interactions clearly show that the usual DR method and the usual HM method should be modified to general forms. After the modifications, the two methods exactly give the same results and the results automatically include the contributions from the seagull interaction, the meson-exchange effect, the contact interactions and the off-shell effect in a correct way. The physical reason why they are exactly the same is also discussed. The expressions in the modified DR method and HM method in the leading order are expressed by  Eq. (\ref{Eq:DR-final}) and Eq. (\ref{Eq:relation-HM-DR}). One should use them to analyze the corresponding experimental data sets and to extract the physical quantities. The new DRs have two additional parameters and this makes it more difficult to extract the full TPE contributions from the experimental data sets. In this work our aim is to show the exact relations between the HM method and DR method, so we do no do the fitting with the experimental data sets.

This work is supported in part
by the National Natural Science Foundation of China under Grant No. 11375044 and No.
11975075. H.Q.Z. would like to thank Shin Nan Yang for his helpful discussion in TPE effect and thank Hiren. H. Pate for his kind help in PackageX. He would like to acknowledge the support of
the National Center for Theoretical Science of the National
Science Council of the Republic of China for his visits in July, 2019.
He also greatly appreciates the warm hospitality extended to him by the Physics Department of the
National Taiwan University during the visits.

\section{Appendix}
In this appendix, the expressions for  $A^{(a,b)}_i(\nu,t)$ are presented.

The expressions for $A^{(a,b)}_{i}$ are a little complex and for simplicity we separate them into two parts : the first part $A^{\textrm{\uppercase\expandafter{\romannumeral1}}(a,b)}_{i}$ comes from the finite parts of the pure loop integrations and the finite trace of Dirac matrix, the second part $A^{\textrm{\uppercase\expandafter{\romannumeral2}}(a,b)}_{i}$ comes from the divergent parts of the pure loop integrations and the trace of Dirac matrix with factor $(d-4)$. Finally we have the following expressions:
\begin{eqnarray}
A^{\textrm{\uppercase\expandafter{\romannumeral1}}(a)}_{1} &=& -\alpha_e^2 \kappa^2\frac{(4M_N^2-t)(2\nu-3t)}{8M_N^3}(\nu^2-\nu_s^2),\nonumber\\
A^{\textrm{\uppercase\expandafter{\romannumeral1}}(a)}_{2} &=& \alpha_e^2 \kappa^2\frac{2\nu-3t}{4M_N}(\nu^2-\nu_s^2),\nonumber\\
A^{\textrm{\uppercase\expandafter{\romannumeral1}}(a)}_{3} &=& -\alpha_e^2 \kappa^2\frac{(8M_N^2-2t+3\nu)t}{8M_N^3}(\nu^2-\nu_s^2),
\end{eqnarray}
and
\begin{eqnarray}
A^{\textrm{\uppercase\expandafter{\romannumeral2}}(a)}_{1} &=& -\alpha_e^2 \kappa^2\frac{4M_N^2-t}{8M_N^3}
\Big[2t(4M_N^2-t)(7t+10\nu)-(11t+4\nu)(\nu^2-\nu_s^2) \Big],\nonumber\\
A^{\textrm{\uppercase\expandafter{\romannumeral2}}(a)}_{2} &=& \alpha_e^2 \kappa^2\frac{1}{4M_N}
\Big[2t(4M_N^2-t)(7t+10\nu)-(11t+4\nu)(\nu^2-\nu_s^2) \Big],\nonumber\\
A^{\textrm{\uppercase\expandafter{\romannumeral2}}(a)}_{3} &=& \alpha_e^2 \kappa^2\frac{t}{4M_N^3}
\Big[-t(4M_N^2-t)(40M_N^2-10t-7\nu)+(28M_N^2-7t-2\nu)(\nu^2-\nu_s^2) \Big].\nonumber\\
\end{eqnarray}
The corresponding expressions for $A^{(b)}_{i}$ can be got by the relations between ${\cal F}_i^{(a)}(t,\nu)$ and ${\cal F}_i^{(b)}(t,\nu)$. These expressions are also checked by the numerical calculation.

\end{document}